\documentclass[preprint,english,showpacs,11pt,floatfix]{revtex4}

\usepackage{babel,amsmath,amssymb,dcolumn}
\usepackage[dvips]{graphics}

\begin{document}

\title{Constraints on the Dark Energy from the holographic
connection to the small \emph{l} CMB Suppression}

\author{Jianyong Shen, Bin Wang}
\email{wangb@fudan.edu.cn} \affiliation{Department of Physics,
Fudan University, Shanghai 200433, People's Republic of China }

\author{Elcio Abdalla}
\email{eabdalla@fma.if.usp.br} \affiliation{Instituto de Fisica,
Universidade de Sao Paulo, C.P.66.318, CEP 05315-970, Sao Paulo,
Brazil}

\author{Ru-Keng Su}
\email{rksu@fudan.ac.cn} \affiliation{China Center of Advanced
Science and Technology (World Laboratory), P.B.Box 8730, Beijing
100080, People's Republic of China
\\Department of Physics, Fudan University, Shanghai 200433,
People's Republic of China }

\begin{abstract}
Using the recently obtained holographic cosmic duality, we reached
a reasonable quantitative agreement between predictions of the
Cosmic Microwave Background Radiation at small \emph{l} and the
WMAP observations,  showing the power of the holographic idea. We
also got constraints on the dark energy and its behaviour as a
function of the redshift upon relating it to the  small \emph{l}
CMB spectrum. For a redshift independent dark energy, our
constraint is consistent with the supernova results, which again
shows the correctness of the cosmic duality prescription. We have
also extended our study to the redshift dependence of the dark
energy.
\end{abstract}

\pacs{98.80.Cq, 98.80.-k}

\maketitle

The latest version of the standard cosmological model describes an
infinite flat universe forever expanding under the pressure of
dark energy. The first year data from the Wilkinson Microwave
Anisotropy Probe (WMAP) manifests a striking agreement with this
model \cite{s1}. However, as originally discovered by the COBE
satellite project,  WMAP results imply a suppression of the Cosmic
Microwave Background Radiation (CMB) anisotropy power on the
largest angular scale as compared to the standard model
prediction. Researchers are now seeking an explanation of such a
wide-angle missing power in the CMB observation
\cite{s2}-\cite{s5}.

Recently, an intriguing attempt to relate the suppression of CMB power in low
multipoles to the holographic idea was put forward \cite{s6}. The
holographic reasoning emerged first in the context of black holes \cite{s7}
and later got extended to the cosmological setting \cite{s8}, having
attracted a lot of attention in the past decade \cite{s9}-\cite{s13}.
It is viewed as a real conceptual change in our thinking about
gravity \cite{s14}. There are many applications of  holography to the
study of cosmology, such as the question of the cosmological constant
\cite{s15}, selecting physical models in inhomogeneous cosmology
\cite{s12}, putting an upper bound on the number of e-foldings
in inflation \cite{s16} and investigating questions related to
dark energy \cite{s17}\cite{s18}. A holographic interpretation of the
features concerned with low multipoles in the CMB power
spectrum \cite{s6} is one more example of how holography can be a useful
tool in understanding cosmology.

A finite universe could be the consequence of a holographic constraint,
giving rise to an effective IR cutoff. In \cite{s6}, the relation between
the features at low multipoles in the CMB power spectrum and the equation
of state of the dark energy was built through a cosmic IR/UV
duality between a global infrared cutoff and the ultra violet cutoff. In
such a cosmic duality model, the qualitative low \emph{l} CMB
features could be well described.

\enlargethispage{\baselineskip}

In this paper we employ the disclosed cosmic duality
to study the nature of the dark energy from the WMAP experimental
data for the small \emph{l} CMB power spectrum. For a redshift independent
equation of state for the dark energy, the result revealed from the
correlation to the observational power spectrum supports the
holographic dark energy model \cite{s17}. We will also investigate
the evolution of dark energy  from the CMB observation in the low multipoles.

Starting from the holographic idea relating the UV and IR cutoffs
as suggested in \cite{s19}, the dark energy density in a flat
universe is \cite{s17}
\begin{equation}\label{e1}
\rho_{\Lambda}=3c^2{M_p}^2L^2\quad ,
\end{equation}
where $L$ is the IR cutoff, $c$ is a free parameter satisfying
$c\geq 1$ as a consequence of the second law of thermodynamics \cite{s17}.
Using definitions $\Omega_{\Lambda}=\rho_{\Lambda}/\rho_{cr}$ with
$\rho_{cr}=3 M_p^2 H^2$, we have at the present epoch the IR
cutoff $ L = \frac{c}{{\sqrt {\Omega _\Lambda ^0 } H_0 }}$.
Interpreting the IR cutoff $L$ as a cutoff of the physical
wavelength, we have $\lambda_c=2L$ \cite{s6}. Thus the smallest
wave number at present is $k_c=\frac{\pi}{{c}}\sqrt {\Omega
_\Lambda ^0 } H_0 $.

The IR cutoff would show up in the CMB angular spectrum in the
Sachs-Wolfe effect,
\begin{equation}\label{e2}
C_l  = \frac{{2\pi }}{{25}}\int_{k_c }^\infty  {\frac{{dk}}{k}}
j_l ^2 (k(\eta _0  - \eta _* ))P_R (k),
\end{equation}
where $P_R=A k^{n_s-1}$ is the curvature power spectrum in the
flat universe, $A$ is the amplitude, $j_l$ is the Bessel function
and $\eta_0-\eta_*$ is the comoving distance to the last
scattering which follows from the definition of comoving time,
\begin{equation}\label{e3}
\eta _0  - \eta _*  = \int_0^{z_* } {\frac{{dz'}}{{H(z')}}}\quad .
\end{equation}

\begin{figure}
\resizebox{0.5\linewidth}{!}{\includegraphics*{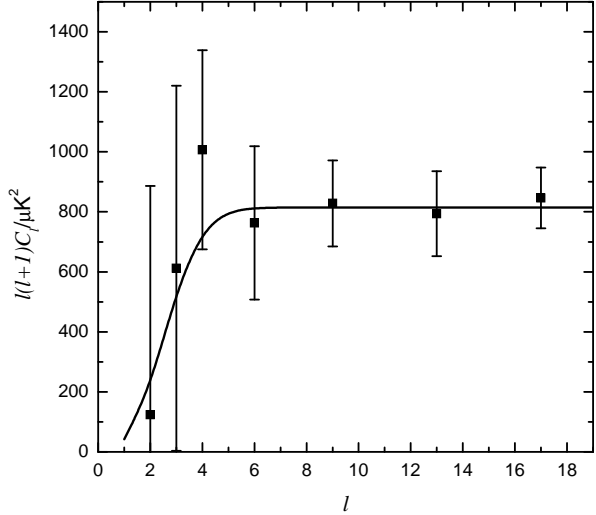}}
\caption{The angular power spectrum at low \emph{l}. The solid
curve stands for the best fit and the
square points for the experimental data from WMAP.}
\label{f1}
\end{figure}

Assuming that the dominant components of energy in the universe
are dark energy and matter, then
\begin{equation}\label{e4}
H^2(z) = H^2_0 [(1 - \Omega _\Lambda ^0 )(1 + z)^3  + \Omega _\Lambda
^0 f(z)]\quad ,
\end{equation}
where $f(z) = exp\{ 3\int_0^z {\frac{{1 + \omega (z')}}{{1 +
z'}}dz'} \}$ \cite{s20}. For a redshift independent equation of state
$\omega=const.$, $f(z)=(1+z)^{3(1+\omega)}$. The distance to the
last scattering depends on $\omega$, which on its turn enters
the Sachs-Wolfe effect expression (\ref{e2}). Thus the relative position of
the cutoff in the CMB spectrum depends on the equation of state of
dark energy. This exhibits the CMB/Dark Energy cosmic duality
first realized in \cite{s6}, that is, there is a correlation between CMB
and the form of the equation of state for the dark energy. We now will
use such a duality to get information on the dark energy from the
observed features about the small \emph{l} CMB spectrum.

Our first attempt is to focus on the case of a redshift independent
equation of state of the dark energy. Using Eqs.(\ref{e2})-(\ref{e4})
to fit the WMAP observational data at low \emph{l} multipoles
\cite{s1}, we have three parameters, namely $(c, \omega, A)$ to be
determined. The best
fitting result of these parameters are determined by the minimum
value of $ \sigma = \sum\limits_i {[y(l_i ) - y_i ]^2 }$, where
$y(l_i)= l_i(l_i+1)C_{l_i} $ is computed by Eq.(\ref{e2}) and
$y_i$ is the observational data from WMAP at different multipoles.
In this numerical analysis, we have taken $z_*=1100$,
$\Omega_\Lambda^0=0.7$ and $n_s=1$. The fitting result is shown in
Fig.\ref{f1} with the position of the cutoff $l_c \sim 4$. It
describes the low \emph{l} CMB features extremely well. The
quantitative agreement with the WMAP observation gives further
support to the holographic cosmic duality.

The best fitting result determines $c=2.1$. The value of $c$ bigger than
unit obtained from the small \emph{l} CMB data fitting gives further
support to the thermodynamical argumentation discussed in \cite{s17}.

The equation of state of dark energy $\omega$ is determined from
the fitting to WMAP data at low multipoles by using the cosmic
duality. By minimizing $\sigma$, we find $\omega=-0.7$ with
minimum value $\sigma_{min}$. In the vicinity of $\omega=-0.7$, we
observed that differences of corresponding $\sigma$'s are small.
Imposing the criterion $ \frac{{\sigma  - \sigma _{\min }
}}{{\sigma _{\min } }} < 10\% $, we obtained the range  of $\omega
\in [-1.3,-0.6]$, which agrees to the result from
supernova\cite{s25}. The more precise observational results on the
exact location of the cutoff $l_c$ and its CMB data are crucial to
constrain the static equation of state of dark energy.

In \cite{s17}, by identifying the IR cutoff exactly with the
future and horizon, the holographic dark energy equation of state
is expressed as
\begin{equation}\label{e4.1}
\omega  =  - \frac{1}{3} - \frac{{2\sqrt {\Omega _\Lambda  }
}}{{3c}}\quad .
\end{equation}
Since $c\geq 1$, the phantom case ($\omega < -1$ corresponding to
$c < 1$) is excluded. Choosing $\Omega_\Lambda=0.7$ and $c=2.1$,
Eq.(\ref{e4.1}) tells us $\omega \approx -0.6$, which is on the
border of the range of the fitting result above.

\begin{figure}[ht]
\vspace*{0cm}
\begin{minipage}{0.5\textwidth}
\resizebox{1\linewidth}{!}{\includegraphics*{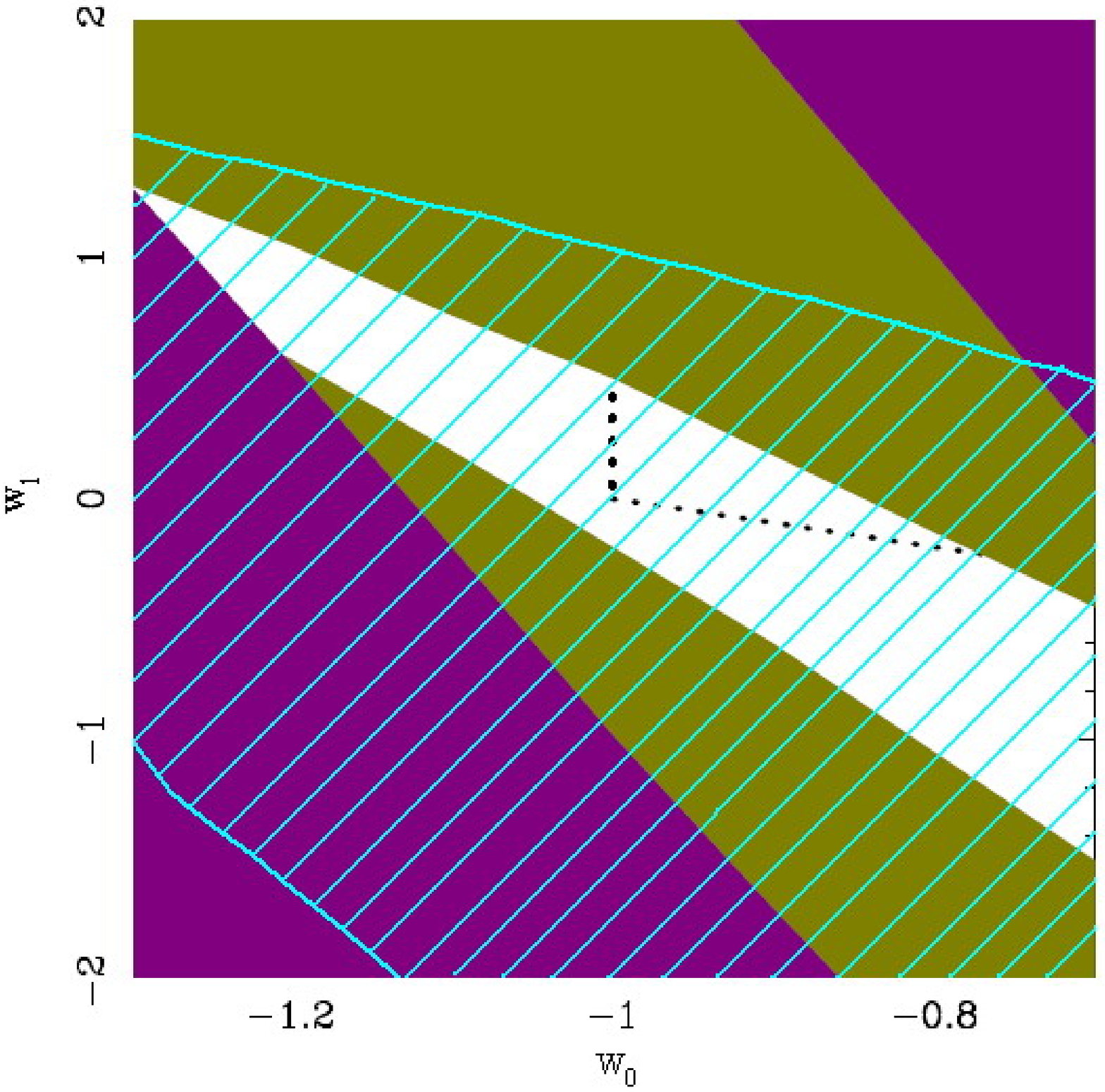}}
\nonumber
\end{minipage}\nonumber
\begin{minipage}{0.5\textwidth}
\vspace*{0.4cm}
\resizebox{1\linewidth}{!}{\includegraphics*{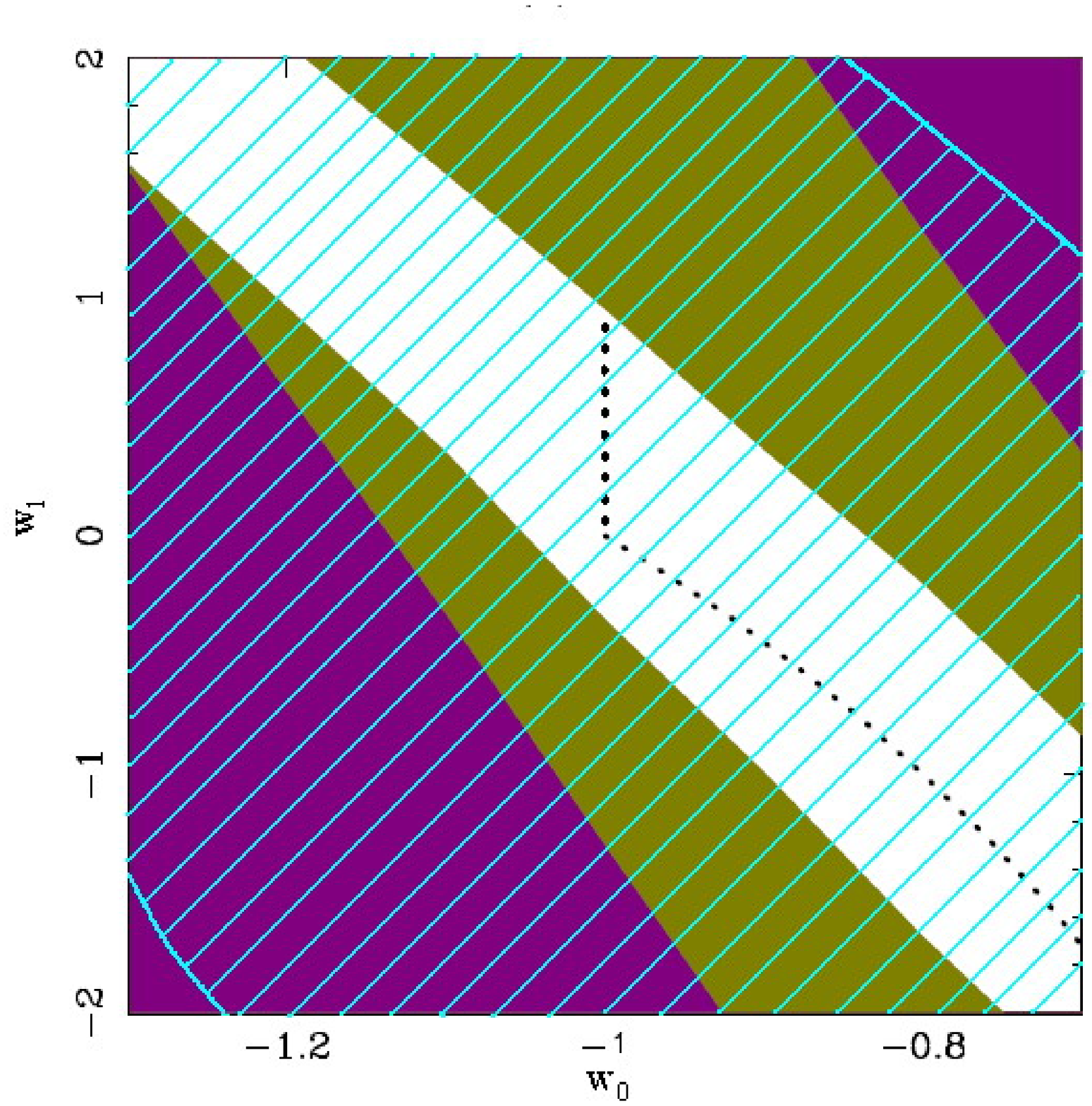}}
\nonumber
\end{minipage} \nonumber
\caption{{The relations between $\omega_1$ and $\omega_0$ for the
model  $\omega^{I}(z)$, Eq.(\ref{e5}), is on the left panel and
for $\omega^{II}(z)$, Eq.(\ref{e6}), on the right panel. The purple area
is ruled out by Supernova data and the yellow ruled out by
WMAP\cite{s23}. The shaded area is the holographic constraint from the
cosmic duality with the suppression position locates in the interval  $4.2
\geq l_c \geq 3.8$.}}\label{f2}
\end{figure}

In the study of the cosmic duality model, we see that the IR
cutoff plays an important role. However, even if an IR/UV duality
is at work in the theory at some fundamental level, the IR cutoff
can not be simply related to the exact future event horizon. Suppose
the IR cutoff has the scale $L=f R_h$, where $R_h$ is the future
event horizon $ R_h  = a\int_t^\infty  {\frac{{dt'}}{{a(t')}}}$
and $f$ is just a constant, then the holographic dark energy
equation of state
\begin{equation} \label{e4.2}
\omega  =  - \frac{1}{3} - \frac{{2\sqrt {\Omega _\Lambda  }
}}{{3(c/f)}}.
\end{equation}
With the fitting result $c=2.1$, Eq.(\ref{e4.2}) allows accommodating
$\omega<-1$ case if $f>2.5$. If $f=1$, $\omega \approx 0.6$, while
$\omega=-1.3$ for $f=3.6$. We expect that future accurate CMB data at
low multipoles can exactly constrain the $\omega$ value and in turn
help us answering the question of whether the effective IR  regulator
is of the same magnitude as the measure of the future event horizon.

\begin{figure}[ht]
\vspace*{0cm}
\begin{minipage}{0.5\textwidth}
\resizebox{1\linewidth}{!}{\includegraphics*{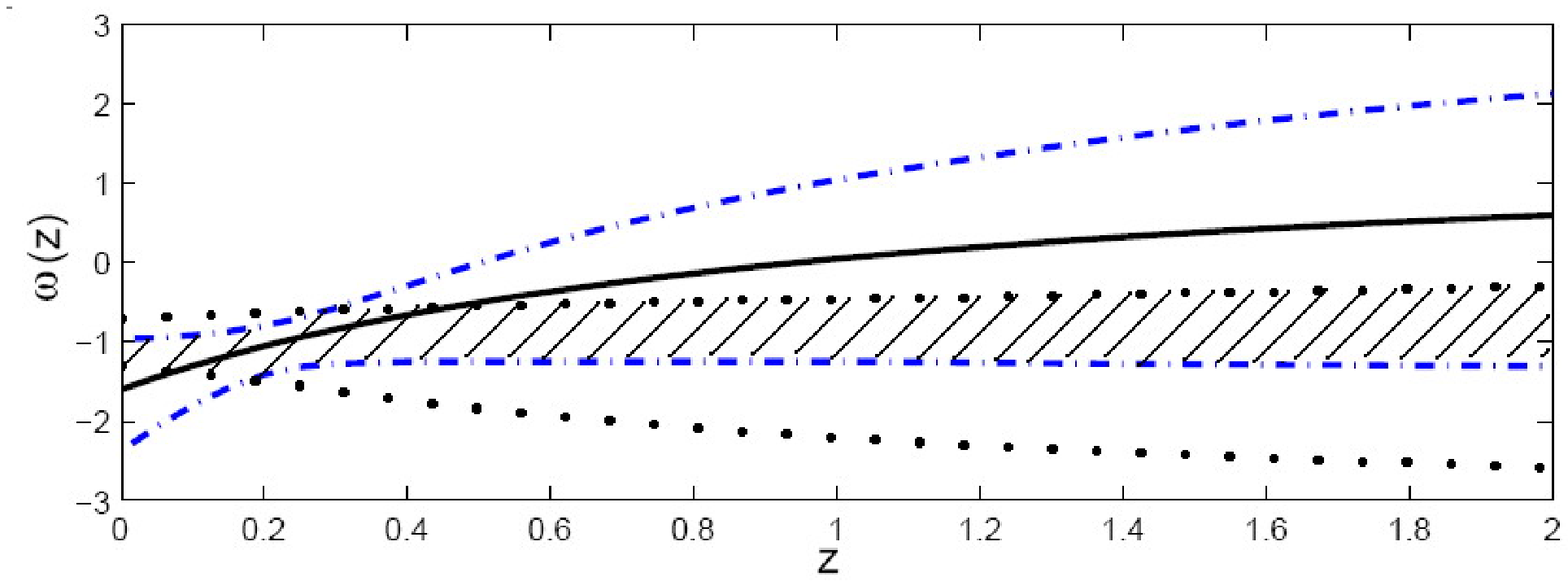}}
\nonumber
\end{minipage}\nonumber
\begin{minipage}{0.5\textwidth}
\vspace*{0.2cm}
\resizebox{1\linewidth}{!}{\includegraphics*{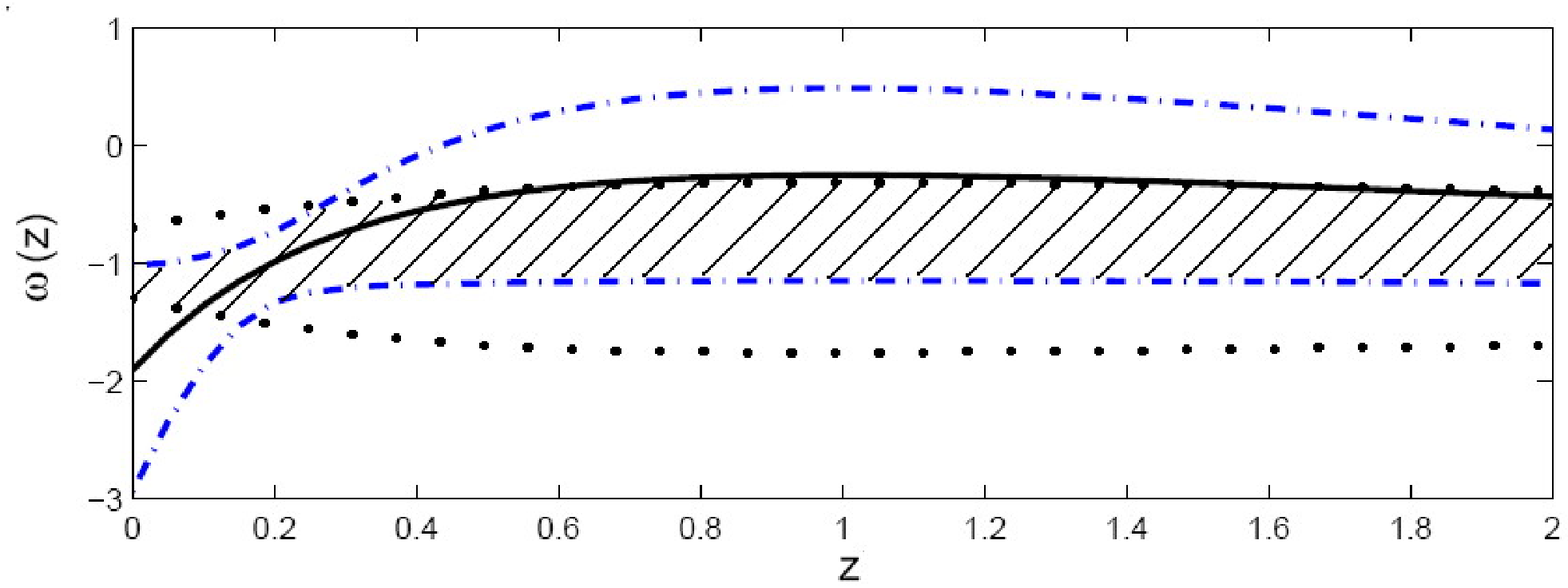}}
\nonumber
\end{minipage} \nonumber
\caption{{The evolution of the models of $\omega^{I}(z)$, Eq.(\ref{e5}),
is on the left panel and that of $\omega^{II}(z)$, Eq.(\ref{e6}),
is on the right panel. The blue dot-dashed lines give the constraints
of their evolutions from the supernova observation. The solid black lines
stand for the evolutions of golden sets of supernova data\cite{s24}.
The black dotted lines are obtained from the low \emph{l} CMB data. The
shaded areas are the overlap between supernova and low \emph{l} CMB
observational constraints.}}\label{f3}
\end{figure}

We now discuss the redshift dependence of $\omega$. We
employ three parameterization discussed
previously \cite{s21}-\cite{s25}. The first two are
\begin{equation}\label{e5}
\omega^I(z)  = \omega _0  + \omega _1 \frac{z}{{1 + z}}
\end{equation}
and
\begin{equation}\label{e6}
\omega^{II}(z)  = \omega _0  + \omega _1 \frac{z}{{(1 + z)^2 }}.
\end{equation}
For both cases we have $\omega(0)=\omega_0$,
$\omega'(0)=\omega_1$. However, the high redshift behaviours of these
functions are different: $\omega(\infty)=\omega_0+\omega_1$ for
Eq.(\ref{e5}) while $\omega(\infty)=\omega_0$ for Eq.(\ref{e6}).
Hence Eq.(\ref{e6}) can model a dark energy component which has
the same equation of state at the present epoch and at high
redshift, with rapid variation at low $z$. For Eq.(\ref{e5}), we
can trust the results only if $\omega_0+\omega_1$ is well below
zero at the time of decoupling.

The third parametrization we will use for the dark energy is
called the Taylor expansion model \cite{s25}
\begin{equation}\label{e7}
\omega^{III} (z) = \frac{{A_1 (1 + z) + 2A_2 (1 + z)^2 }}{{3[A_0 +
A_1 (1 + z) + A_2 (1 + z)^2 ]}} - 1\quad .
\end{equation}
The effect of low-\emph{l} CMB suppression can provide constraints on
dynamical models of dark energy. Using Eqs.(\ref{e2}-\ref{e4}) we
can obtain the behavior of the variation of $\omega$ from the CMB
data through cosmic duality. The exact location of the cutoff
$l_c$ and the shape of the spectrum at low \emph{l} are crucial to
investigate the redshift dependence of $\omega$. Here we narrow
the position of the cutoff $l_c$ in the multipole space in the
interval $3.8<l_c<4.2$. The free parameter $c$ is set to be $2.1$.
\begin{figure}
\resizebox{0.5\linewidth}{!}{\includegraphics*{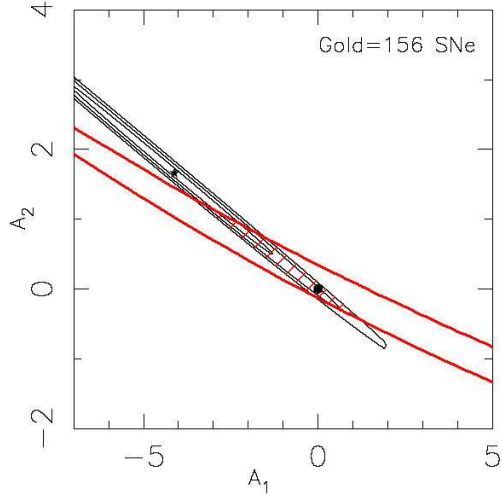}}
\caption{The constraint on the parameters $A_1$ and $A_2$ for the
model Eq.(\ref{e7}). The area enclosed by the black curves is
allowed by supernova data\cite{s25}. The red shaded area is the
overlap of constraints from supernova and small $l$ CMB observations.}
\label{f4}
\end{figure}

Fig.\ref{f2} exhibits the constraints on the redshift dependence
of $\omega$ described by the first (left) and second (right)
parameterizations, respectively. The regions blancked out in purple
are ruled out by supernova constraints and the green regions are ruled out
by WMAP \cite{s23}. Shaded areas are holographic constraints we
obtained from the small \emph{l} CMB spectrum. They have overlaps
with the constraints gotten in \cite{s23}.

With the more precise values for the supression region and a related
better shape of the spectrum for low-\emph{l}, the shaded area will be
reduced and the holographic constraints from the cosmic duality
will become tighter.

\begin{figure}
\resizebox{0.5\linewidth}{!}{\includegraphics*{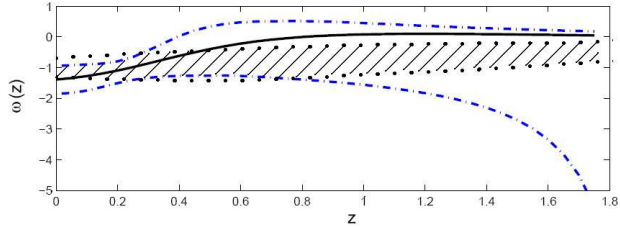}}
\caption{The evolution behavior of the parametrization Eq.(\ref{e7}). The
area between the blue dot-dashed lines are the constraints from
supernova data. The solid black line stands for the evolution of the
golden set \cite{s24}. The black dotted lines are obtained from small
\emph{l} CMB observation. The shaded area is the overlap of supernova and
low \emph{l} CMB constraints.}
\label{f5}
\end{figure}

In order to get a better insight, we have shown in Fig.\ref{f3} the allowed
values of $\omega(z)$ as a function of redshift $z$ with supernova
data \cite{s24} (blue dotted dash lines) and the small \emph{l} CMB
data (black dotted lines). The solid black lines stand for the
evolution of the golden set of supernova data \cite{s24}. The left
panel shows the result for the first parametrization. In contrast with
the supernovae results, the correlation between the dark energy
and small \emph{l} CMB spectrum puts $\omega_1+\omega_0$ well
below zero. Thus the dark energy density did not dominate over the
matter density at high $z$, from CMB observation. The shaded area
is the combined constraint from the supernovae and small \emph{l}
CMB data. Including the low-\emph{l} CMB constraint, the first
parameterization cannot be ruled out as a choice of modelling
dynamic dark energy even at high $z$. The right panel in
Fig.\ref{f3} exhibits the behavior of $\omega^{II}(z)$ for the
second parametrization. It is clear from both the supernovae and
CMB constraints that the matter density did dominate over the dark
energy density at high $z$ in this parametrization.

We now investigate the third Ansatz for the redshift dark energy dependence.
Fig.\ref{f4} shows the relation of $A_2$ and $A_1$. The area
enclosed by the black curves is allowed by supernova
observation\cite{s25}. The red shaded area is the overlap of the
combined supernova and small \emph{l} CMB suppression where the
cutoff position is within the interval $3.8\leq l_c \leq 4.2$. It
leads to tighter constraints on the model. The behavior
of the evolution of such a dark energy Ansatz is shown in
Fig.\ref{f5}, where the blue dotted dash lines stem from the
supernova data \cite{s24} and black dotted lines from
low-\emph{l} CMB data. The solid line stands for the evolution of
the golden set of supernova data. It is clear that using the
cosmic duality scenario the small \emph{l} CMB spectrum puts
tighter constraint on the evolution of $\omega$.

In summary, we have employed the holographic cosmic duality to
study the nature of the dark energy. Using the cosmic
IR/UV duality, we obtained the quantitative agreement of low
\emph{l} CMB features to WMAP observation, which shows the effectiveness
of the holohraphic idea. By the correlation
disclosed between the dark energy and small \emph{l} CMB power
spectrum, we have obtained constraints on dark energy models from
the low \emph{l} CMB data. For the static equation of state, we
have got the consistent range of $\omega$ with that from supernova
experiment. This shows again the correctness of the idea of
holographic cosmic duality. We have also studied constraints for
the dynamic dark energy model and compared the results obtained
from supernova data. To obtain more precise constraints on the
dark energy through this holographic cosmic duality and to answer
the question whether IR cutoff is exactly the
future event horizon, exact location of the suppression position
and precise shape of the CMB power are crucial, especially employing
independent methods such as baryonic oscillations in future surveys
\cite{abdraw}.

\begin{acknowledgments}
This work was partially supported by  NNSF of China, Ministry of
Education of China, Ministry of Science and Technology of China under grant NKBRSFG19990754 and Shanghai Education Commission.
E. Abdalla's work was partially supported by FAPESP and CNPQ,
Brazil. R.K. Su's work was partially supported by the National
Basic Research Project of China.
\end{acknowledgments}


\end{document}